\documentstyle[onecolumn]{mn}
\def\yr{{\rm yr}\ }
\def\Msun{M_{\odot}\ }
\def\Mpc{{\rm Mpc}\ }
\def\etal{{\it et al.}\ }

\def\kmsmpc{\hbox{km s}^{-1} \hbox{Mpc}^{-1}}
\def\M10{{\times 10^{10} M_{\odot}\ }}

\begin{document}

\title{Hydrodynamical simulations of galaxy formation:
effects of supernova feedback}

\author[G. Yepes, R. Kates, A. Khokhlov and A. Klypin]
 {G. Yepes,$^1$
 R. Kates,$^2$
 A. Khokhlov,$^3$
and A. Klypin$^4$ \\
$^1$ Departamento de F\'{\i}sica Te\'orica C-XI, Universidad
Aut\'onoma de Madrid, Cantoblanco 28049, Madrid, Spain \\
$^2$ Astrophysikalisches Institut Potsdam, Potsdam, Germany \\
$^3$  Department of Physics and Astronomy, University of
Texas at Austin, USA \\
$^4$ Department of Astronomy, New Mexico State University,
 Las Cruces, NM 88001, USA
}

\maketitle

 \begin{abstract} In this paper we numerically simulate some of the
most critical physical processes in galaxy formation: The supernova
feedback loop, in conjunction with gas dynamic processes and
gravitational condensations, plays a crucial role in determining how
the observable properties of galaxies arise within the context of a
model for large-scale structure.  Our treatment incorporates a
multi-phase model of the interstellar medium and includes the effects
of cooling, heating and metal enrichment by supernovae, and
evaporation of cold clouds. The star formation happens inside the
clouds of cold gas, which are produced via thermal instability. In
this paper we simulate the galaxy formation in standard biased Cold
Dark Matter (CDM) model for a variety of parameters and for several
resolutions.

In our picture, supernova feedback regulates the evolution of the gas
components and star formation.  The efficiency of cold cloud
evaporation by supernova strongly influences star formation rates.
This feedback results in a steady rate of star formation in ``large''
galaxies (mass larger than $(2-3)\times 10^{11}\Msun$ within 100~kpc
radius) at a level of $(1-10)\Msun$ per year for $z < 3$ ($H_0=50$
Km~s$^{-1}$~Mpc$^{-1}$).
Supernova feedback has an even stronger effect on the evolution of ``dwarf''
galaxies.  Most of the dwarf galaxies in our models have a small fraction of
stars and extremely low luminosities:  $M_R>-15$ for parent dark-halo masses
$M_{\rm tot}<(2-3)\times 10^{10}\Msun$ within a 50~kpc radius.

The observational properties (colors, luminosities) of galaxies
identified in the simulations are computed using a stellar population
synthesis model.  In the case of both large and small galaxies, the
distribution of luminous matter (stars) is strongly biased with
respect to the dark matter.  For a range of parameter values and
resolutions we find an approximate biasing measure of the form $
\rho_{\rm lum}=(\rho_{\rm dm}/133)^{1.7}$, for overdensities exceeding
about 1000.  Deviations from this relation depend strongly on the
environment.  For halo masses exceeding $2 \times 10^{10} \Msun$, the
dependence of the absolute visual magnitude $M_V$ on the total mass
can be approximated as $ M_V=-18.5 -4\log ( {M_{\rm
tot}}/{10^{11}\Msun}), $ with a scatter of less than 1/2 magnitude.

\end{abstract}


\section{Introduction}

In the past, an important reason for the attractiveness of dark-matter models
for large-scale structure has been their complete neglect of nongravitational
forces, permitting rather far reaching conclusions to be drawn using relatively
simple physical considerations combined with quite sophisticated mathematical
and statistical methods.  Improvements in computational resources now permit
numerical simulations of large-scale structure to approach a resolution at which
the neglect of non-gravitational effects, rather than the limited numerical
resolution, is the principal source of uncertainties.

In this paper, we study the formation and evolution of galaxies in the context
of large-scale structure.  We focus our attention on the role of supernova
feedback and hydrodynamics in combination with gravitational dynamics in
determining the evolution of some of the most important observational properties
of galaxies, including luminosities and colors, and the extent of their
dependence on mass and environment.

The formation and evolution of galaxies and the dynamics of the intergalactic
medium are complicated processes which cannot be modelled purely from first
principles and are not yet completely understood.  One of the main challenges to
modeling is the fact that a variety of effects over an exceptionally wide range
of scales all come into play:  Cosmological fluctuations on scales larger
than tens of megaparsecs are important because these scales define the
large--scale structure of the galaxy distribution -- positions and sizes of
clusters, superclusters, filaments, and voids.  Many parameters of galaxies
(like the morphological type, rate of collisions and merging) are known to
depend on the position of a galaxy within the structures.  Conditions for galaxy
formation are different for galaxies inside groups and in the field, which
brings in a scale of one megaparsec.  Although the dark matter still dominates
gravitationally on scales above 100~kpc, the baryonic component becomes
more and more important as we descend from the megaparsec scale.  The
dissipation of energy by the hot gas definitely is an important factor on scales
below 100~kpc.  For the formation of the luminous component of galaxies, scales
below 10~kpc are essential.  These scales define the global dynamics of the
galaxy -- density and velocity profiles, spiral arms, and so on.  In turn, as
discussed in more detail below, the global galactic dynamics regulates the
formation of cold gas clouds (in our galaxy -- molecular clouds) inside of which
stars are produced.  The characteristic scale of the clouds is 10--100~pc.  The
clouds themselves have a very complicated internal structure.

Thermal instability plays an important role in these processes:  It has long
been appreciated (Field 1965, Balbus 1986, Nulsen 1986 ) that gas subject to
cooling (and perhaps heating) processes may become thermally unstable.  In
particular, for primordial abundances, pressure equilibrium, and in the absence
of photoionization, it is known that thermal instabilities are possible for gas
within a wide range of temperatures that are quite typical in a cosmological
setting.  For example, thermal instabilities are important in the
theory of globular clusters (Fall \& Rees, 1985).  The situation is more
complicated if photoionization is properly taken into account (M\"ucket \&
Kates, 1996), but the general picture remains that at densities and temperatures
typical for a galactic halo, and with reasonable estimates for the ionizing
flux, instabilities will evolve on a timescale that is sufficiently short
compared to appropriate dynamic timescales such as a free fall time.

One important consequence of thermal instability 
is that it causes the formation of cool clouds, whose temperature
continues to drop until the cool gas becomes neutral and essentially stops
losing energy.  This occurs at a temperature of roughly $10^4 K$.  Now, if the
heat gain due to conductivity is greater than the loss due to radiation, clouds
will not persist.  However, beyond a critical size, clouds tend to grow because
conduction is not sufficient to evaporate them (Begelman \& McKee, 1990).  
When the size of a cloud approaches the
Jeans limit, it starts to form stars with all the ensuing
complications.  Eventually, the cloud presumably stops growing as a
result of reheating due to stellar winds, supernova explosions, and
radiation from young stars.  Subsequent cooling of the hot gas could
form new clouds.

Cold clouds are almost certainly an essential ingredient in the dynamics
of galaxies, particularly in star formation and the interstellar medium.
Their importance has long been appreciated, and there
have been many refinements in the multi-component
picture of the interstellar medium (Oort, 1954; Spitzer, 1965;
Field \& Saslaw, 1965; Field, Goldsmith \& Habing, 1969;
McKee \& Cowie,  1977)  Thermal instabilities 
and cloud formation are important in the
theory of globular clusters (Fall \& Rees, 1985). 

Another important component of galaxy formation and evolution is
supernovae and young stars.  Supernovae eject enormous amounts of
energy into the surrounding gas.  Depending on 
circumstances, this process can either stimulate
or inhibit star formation (McKee \& Ostriker 1977, Lada 1985, Wolf \&
Durisen 1987, Spitzer 1990).  In a cold, dense cloud, it can 
stimulate star formation by increasing the pressure at the periphery
and thus causing collapse in the interior.  On the other hand, close
to the supernova the shock heats the gas and evaporates clouds 
(McKee \& Cowie, 1975). This
eventually turns off star formation.  Supernovae also enrich the gas
with metals, which drastically affects cooling rates and the
conditions for star formation.
Much of the energy of supernovae may contribute to 
the velocity dispersion of cold clouds (McKee \& Ostriker, 1977;
Cowie, et al., 1981).

A proper description of the above phenomena in the context of
galaxy formation would not be complete
without an adequate treatment of gas dynamics.  Hydrodynamic
simulations in a cosmological context were pioneered by Evrard (1988),
Cen \etal (1990), and Katz \& Gunn (1991), but for a comprehensive
discussion of the different numerical methods we refer the reader to
the comparative paper of Kang et al (1994).  The heating of the gas by
supernovae and its role in modulation of galaxy formation have been
studied by a number of authors (Baron \& White 1987, Cen \& Ostriker,
1992; Katz, 1992; Navarro \& White 1993; Kauffmann, White \&
Guiderdoni, 1993; Metzler \& Evrard, 1994; Mihos \& Hernquist, 1994a;
Cole et al 1994).  Navarro \& Steinmetz (1996) studied the effects
of photoionization on angular momentum and cooling of disk galaxies
without including the feedback of energy due to star formation and
supernovae and found that photoionization alone was not adequate
for proper understanding of the "overcooling" phenomenon.  
Taken together, these previous studies tend to confirm the
importance of including a realistic description of the star-gas
interaction loop including the mechanical effects of supernova
explosions and multiple gas phases.  This point will be addressed further
in the conclusions.

This paper is organized  as follows:
In \S 2 we derive the system of equations describing our multiphase
model.  In \S 3, we express the equations in a form suitable for
application of numerical techniques.  For solution of the hydrodynamic
equations, we have used the piecewise parabolic method (PPM; Colella
\& Woodward 1984); the collisionless components are treated by
particle mesh methods (Kates, Kotok, \& Klypin, 1990; Klypin \& Kates,
1991).  The PPM module was written by one of us (A.Kh). Several tests
of the code are reported in \S 4, together with parameter studies
concerning the effects of supernova feedback and enhanced cooling.  \S
5 gives the parameters for a set of CDM simulations and describes
algorithms for finding galaxies and computing their colors and
luminosities using a stellar population synthesis model.  Results are
discussed in \S 6, including trends in luminous matter vs. total mass,
luminosity vs. total mass, colors, and the importance of environmental
effects.  Although the results need to be confirmed by more
simulations, we conclude in \S 7 that the supernova feedback loop in
combination with hydrodynamics and gravitation lead to a rich variety
of dynamical processes in galaxy formation.  These effects act as
``hidden variables'' and contribute significantly to the scatter of
phenomenological relationships between the various observable
properties of galaxies.
We assume $H_0= 50\;\; \kmsmpc$ throughout this paper unless otherwise stated.

\section{Physical assumptions and motivations}

\subsection{Multi-phase medium}

The matter in the simulated Universe consists of four phases. 1) The
dark matter (labeled by a subscript ``dm'') in th form of weakly
interacting collisionless particles is the main contribution to the
mean density of the universe ($\Omega_{\rm dm}=1-\Omega_{\rm b}$). The
baryonic component is described as a medium consisting of three
interacting phases: 2) hot gas (labeled by subscript h, $T_{\rm h}>2
\times 10^4$ K), 3) gas in the form of cold dense clouds (subscript c,
internal temperature $T_{\rm c}~=~10^4$ K) resulting from cooling of
the hot gas, and 4) ``stars'' (subscript $*$), formed inside cold
clouds and treated as collisionless particles.  Thus, the total
density $\rho(\vec r)$ is the sum of four components:
\begin{equation}
   \rho = \rho_{\rm dm} +\rho_{\rm h} +\rho_{\rm c} +\rho_*.   
\end{equation}

This picture is consistent with work on models
of galaxy formation and evolution (see for example
Nulsen 1986; Thomas, 1988a,b; Hensler and Burkert, 1990; Daines et al
1994; Nulsen \& Fabian, 1995)  and in this context appears to be superior
to a treatment of the gas component as a one-phase medium.

\subsection{Cooling of the hot gas} \label{sect:22}

A proper treatment of cooling requires consideration of processes
occurring below the numerical resolution of the code.  Despite certain
simplifications, our picture relies heavily on the comprehensive
theory of the interstellar medium of McKee \& Ostriker (1977), who
provided a detailed description of the physics in dense
regions.  This theory presupposes the existence of cold clouds forming
due to thermal instability in approximate pressure equilibrium with
the surrounding hot gas.  Heat is conducted from the hot to the cold
phase through warm interfaces surrounding the clouds, but sufficiently
large clouds persist.  Despite the existence of multiple phases with
different temperatures, the rate of cooling can still be computed from
knowledge of the mean parameters of the hot gas.

Let us denote the {\it local} cooling rate of the plasma due to
radiative processes by $dE/dt=-\Lambda_r(\rho,T) $, where $\rho$ and
$T$ here represent the {\it true} local values of the gas density and
temperature.  According to the theory of McKee and Ostriker, the rate
of energy loss expressed in terms of {\it average} gas density
$\rho_h$ and temperature $T_h$ will be higher than the nominal rate
$-\Lambda_r(\rho_h,T_h $) by a {\it cooling enhancement factor} $C$:
$dE/dt=-C\Lambda_r(\rho_h,T_h )$ (McKee \& Ostriker used $\beta$ for
this parameter).  In this paper, we take their estimate $C=10$.  In
our numerical code the averaging happens over a resolution cell.  This
value of parameter $C$ takes into account all effects resulting from
unresolved density inhomogeneities including radiation from interfaces
between the cold clouds and the surrounding hot gas.  Cooling depends
also on the chemical composition of the gas.  In order to incorporate
the effects of metallicity into our code, we assume solar abundance in
regions where either thermal instability is present (assuming the gas
density exceeds an appropriate threshold; see below) or previous star
formation has occurred.  Otherwise, the region is considered not to be
enriched by metals, and the cooling rate for a gas of primordial
composition (Fall \& Rees 1985) is assumed.  In \S 2.6, we will
describe our procedure for choosing cooling rates in detail.

The ionization of the intergalactic and interstellar medium by UV
photons emitted by quasars, AGNs, and by nonlinear structures in the
process of formation has two important consequences for our code.
First, we assume that the gas outside dense regions ($\rho_{\rm gas}<
2\langle \rho_{\rm gas} \rangle$, where $\langle\rho_{\rm gas}\rangle$
is the mean cosmological gas density) is ionized and has a temperature
of $T=10^{3.8}$K, consistent with recent results (Giroux \& Shapiro,
1996; Petitjean, M\"ucket, \& Kates, 1996; M\"ucket et al., 1996).
 Secondly, in low to moderate-density regions, heating due to the ionizing
flux suppresses the thermal instability (M\"ucket \& Kates, 1996) 
and thus the formation of cold clouds
(and ultimately stars).  Hence, we allow cold clouds to form
only if $\rho_{\rm gas}/\rho_{\rm cr} > {\cal D}\cdot\Omega_{\rm bar}$, with
${\cal D}\sim 50-100$.  

In addition to the above processes we include the energy loss 
due to Compton cooling, given by
$\Lambda_{\rm
   Comp}=7\times 10^{-36}n_HT_ea^{-4}$, where $n_H$ is the number density
   of hydrogen ions, $T_e$ is the electron temperature, and $a$ is the
expansion parameter.

\subsection{Transfer of gas to the ``cold'' phase}

In this model, there are two mechanisms for gas to leave the hot phase
and enter the cold phase.  First, if the temperature of the hot gas
drops below a threshold temperature $T_{\rm lim}=2\times 10^4$ K, and
if  $\rho_{\rm gas}/\rho_{\rm cr} > {\cal D}\cdot\Omega_{\rm bar}$ we
immediately transfer all the hot gas to the cold phase, thus making it
available for star formation. This process gives the following terms
in the continuity equations for the gas:
\begin{equation}
      \left({d\rho_{\rm h} \over dt}\right)_{{\rm hot}\rightarrow {\rm cold}}
      = -\frac{\rho_{\rm h}} {t_{\rm cool}},
   \hskip 1.5em t_{\rm cool} = {\epsilon_{\rm h} \over 
                                   \partial \epsilon_{\rm h}/ \partial t},
    \hskip 1.5em \frac{\rho_{\rm gas}}{\rho_{\rm cr}}> {\cal D}\cdot\Omega_{\rm bar}    \hskip 1.5em T < T_{\rm lim},
\end{equation}

where $t_{\rm cool}$ is the cooling time, and $\epsilon_{\rm h}$ is
the thermal energy per unit mass.
The transfer may be treated as immediate if
the cooling time is very short compared to 
a timestep. This typically happens at  $T\approx 2\times 10^4$ K.
A second way of transferring gas to the cold phase is by sufficiently rapid
thermal instability.
The temperature range for creation of cold clouds depends in reality
on the ionizing flux and the local density (M\"ucket \& Kates, 1996).
Here, we model these restrictions by checking the condition
$T<T_{\rm inst}$, where in this paper 
$T_{\rm inst} = 2\times 10^5$ K.    
To estimate the rate of growth of mass in cold clouds, we assume that
the energy emitted by the hot gas was actually lost by the
gas, which was hot and became cold.  
(Note that according to McKee \& Ostriker (1977), the filling factor of the cold clouds
is small.) If $\epsilon_{\rm h}$ and
$\epsilon_{\rm c}$ are the internal thermal energies
 per unit mass of hot and cold gas, then the change of energy
 of the system due to radiative cooling is 
\begin{equation}
\label{eq:EQA}
\left( \frac {d(\rho_{\rm h} \epsilon_{\rm h} 
+\rho_{\rm c} \epsilon_{\rm c})}{dt}\right)_{\rm cooling}
   =        -C\Lambda_r(\rho_h, T_h),
\end{equation}
where $C$ is the enhanced cooling factor due to unresolved clumpiness.
We take $\epsilon_{\rm c}= {\rm constant}$, which means that the cold gas
cannot cool below $T_{\rm lim}$.  For an ideal gas, these assumptions
imply extra terms in continuity equations:
\begin{equation}
\label{eq:EQB}
\left({d \rho_{\rm h} \over dt}\right)_{\rm therm.inst}
   = -\left({d\rho_{\rm c} \over dt}\right)_{\rm therm.inst} = -{C
   \Lambda_r(\rho_h, T_h) \over \gamma \epsilon_{\rm h}
   -\epsilon_{\rm c}}, 
\end{equation}
where $\gamma$ is the ratio of specific heats.

Because of the frequent exchange of mass between the hot and cold gas
phases, it is reasonable to consider the hot gas and cold clouds as {\it one}
fluid with rather complicated chemical reactions going on within it.  
Thus, we follow the motion of the {\it hot} component and integrate
the change of the total density of the gas (index
$gas$):
\begin{equation}
   \rho_{\rm gas}=\rho_{\rm c} +\rho_{\rm h}. 
\end{equation}
The cold and hot gas are assumed to be in pressure equilibrium locally. So,
$P_{\rm gas}=P_{\rm c}=P_{\rm h}\equiv (\gamma-1)u_{\rm h},$
$u_{\rm h}=\rho_{\rm h} \epsilon_{\rm h}$, where 
$u$ and $\epsilon$
 are internal energies per unit volume and per unit
 mass correspondingly.

The velocity associated with the fluid involves an averaging by the
code over a cell size.  We imagine that, in a multiphase medium with
supernovae, the hot gas is in reality ``windy'' at scales below the
cell size and similarly that the cold clouds have a significant
velocity dispersion (McKee \& Ostriker, 1977; Cowie, et al., 1981;
Hensler and Burkert, 1990.)  We therefore associate an ``effective"
temperature with the existence of the cloud component which in the
present paper is taken simply equal to the ambient hot gas
temperature, i.e., there is a pressure $P_{\rm disp}\propto \rho_cT_h$
associated with the cold gas component.  Although this pressure may
represent an overestimate, the overestimate in the {\it total}
pressure (hot plus cold) would only be significant if the cold gas
fraction could exceed the hot gas fraction.  However, in our model
this situation never arises due to the heating by supernovae as
explained below.

\subsection{Star formation}

 Star formation, which is assumed to occur only in cold clouds, leads to
 a decrease in the  density of the cold component given by
 \begin{equation}
 \left(\frac{d\rho_{\rm c}}{dt}\right)_{\rm star-formation}
 = -\frac{\rho_{\rm c}}{t_*},
 \end{equation}
  where we take a fixed characteristic star formation
 time constant $t_*\approx 10^8~\yr$.
 The actual time-scale for
 star formation can exceed $t_*$ and will depend strongly
 on the rate of the conversion of hot gas to the cold phase.  
 The lifetime of massive stars is about $10^7\yr$ or even shorter;
 thus most of stars with $M_*$ larger than $(10-20)\Msun$
 will explode as supernovae during one timestep.
 Stars in the mass interval from (5--7)$\Msun$ to 10$\Msun$
 will explode on a longer time-scale, but they
 produce less energy, and therefore in view of the 
 uncertainties in supernova energy, the energy input
 due to these stars is
 not included here. Due to the explosion of massive stars, the rate of
 growth of mass in the form of long-lived stars is decreased by a fraction
   $\beta$: 
\begin{equation} 
 \label{eq:srate} 
  \frac{d\rho_*}{dt}=\frac{(1-\beta)\rho_{\rm c}}{t_*}. 
\end{equation}
The precise value  of  $\beta$ is sensitive to to the form
of the initial mass function (IMF), particularly its low-mass limit.
The principal source of uncertainty is that this low fraction of mass in
massive stars relies on the {\it present} IMF, whereas we would like to
apply it to very early epochs of galaxy formation.  The main difference
is in the abundance of heavy elements.  It is possible that in a
hydrogen-helium plasma, only supermassive stars could be formed.  At the
same time, enrichment of the medium could proceed very fast, changing
the chemical composition of gas and, as the result, the IMF.  In view of
these uncertainties, we assume here that $\beta$ is constant in time,
while keeping in mind that other assumptions are permitted and might
lead to different results. Assuming a Salpeter IMF, we estimate the
fraction of mass in stars with mass larger than $10\Msun$ to be
$\beta=0.12$. 
  
\subsection{Effects of supernovae}
Evaporation of cold clouds is an important effect of supernovae on the
interstellar medi\-um (McKee \& Ostriker 1977, Lada 1985).  We
incorporate this effect by supposing that the total mass of cold gas
heated and transferred back to the hot gas phase is a factor $A$ higher
than that of the supernova itself.  We will refer to $A$ below as
the {\it supernova feedback parameter}.  Realistic evaporation is a much
more complicated process.  The rate of evaporation could depend on, among
other things, the energy of the supernovae, the cloud spectrum, and
the ambient density.  McKee and Ostriker (1977) give an estimate of the
evaporated mass which scales with the energy of the supernova $E$ and
the gas density as $E^{6/5}\cdot n_{\rm h}^{-4/5}$.  However, at present we
do not include the
dependence on the gas density, because here $n_{\rm h}$ refers to the
local density, 
which is relevant for the propagation of the supernova shock, whereas
our $n_{\rm h}$ is the mean density of the hot gas averaged over quite
large volume.

Assuming that the energy input due to supernovae is proportional to the
total mass of supernovae, we obtain for the rate of mass exchange
due to evaporation 
\begin{equation}
\left(\frac{d\rho_h}{dt}\right)_{\rm evap}=-\left(\frac{d\rho_c}{dt}
\right)_{\rm evap}= \frac{A\beta \rho_{\rm c}}{t_*}
\end{equation}
Correspondingly,
the net energy supplied to the hot gas phase by supernovae is:
\begin{equation}\label{eq:super}
\left(\frac{d\rho_h\epsilon_h}{dt}\right)_{\rm SN} 
   =  \frac{\beta\rho_{\rm c}}{ t_*}
      [\epsilon_{\rm SN}+A \epsilon_{\rm c}] .
\end{equation}
Here $\epsilon_{\rm SN}=$ $\langle E_{\rm SN}\rangle/\langle M_{\rm
SN}\rangle$, where for the supernova energy and mass we take the
values $\langle E_{\rm SN}\rangle =10^{51}{\rm erg}$, $\langle
M_{SN}\rangle = 22\Msun$, respectively (Salpeter IMF).  Admissible
values of the supernova feedback parameter $A$ are constrained by the
condition that the energy of a supernova explosion be larger than the
energy required for the evaporation of cold clouds.  This restriction
leads to $\epsilon_{\rm SN} > A\cdot (\epsilon_{\rm h}-\epsilon_{\rm
c})$.  For the above supernova energy and mass and for hot gas
temperature $T_{\rm h} \approx 10^6$ K, this condition implies the
restriction $1 \le A < 250$.  We will discuss implications of varying
the supernova feedback parameter $A$ and more detailed evidence
constraining its value in \S \ref{tests}.

Of course, $A$ could depend both explicitly on time and on local
chemical abundances.  In the absence of a model for these effects, we
take constant $A$, which nevertheless includes the main effects and
{\it together} with the other effects (energy input due to supernovae
and radiative cooling) gives a rather complicated nonlinear picture of
star formation.  (See e.g.  Thomas, 1988a; Navarro \& White 1993 for a
different way of modelling the SN feedback).

\subsection{Model equations}\label{s:modeleq}

Expressed in expanding coordinates ($a$ is the expansion parameter),
the system of equations to be solved numerically thus comprises:
\begin{enumerate}

\item Equations for the dark matter component (collisionless particles):
\begin{equation}
 \frac{d{\bf p}}{dt} =-\nabla\phi, \qquad \frac{d{\bf x}}{dt}=\frac{{\bf p}}{a^2},
\end{equation}
where ${\bf p}=a^2\dot{\bf x}$ and $\bf x$ are the momentum and the
position of a dark matter particle, respectively (Peebles 1980).

\item Equations for the ``stars'' (collisionless particles) with a
source term due to the production of new stars in cold clouds:
\begin{equation}
   \frac{d{\bf p}_*}{dt} =-\nabla\phi, \qquad \frac{d{\bf x}_*}{dt}=\frac{{\bf p}_*}{a^2}, 
\end{equation}
\begin{equation}
   {\partial \rho_* \over \partial t} 
      +3\left({\dot a \over a}\right)\rho_* +
            \frac{1}{a}\nabla \rho_* \vec v_* =
               {(1-\beta)\rho_{\rm c} \over t_*}, 
\end{equation}

\item The continuity equation for the gas:
\begin{equation}
{\partial \rho_{\rm gas} \over \partial t}
    +3\left({\dot a \over a}\right)\rho_{\rm gas} +
      \frac{1}{a}\nabla \rho_{\rm gas} \vec v_{\rm gas} =
         - {(1-\beta)\rho_{\rm c} \over t_*},
\end{equation}

\item The system of hydrodynamical equations for the ``hot'' component:
\begin{eqnarray}
   {\partial \rho_{\rm h} \over \partial t} 
      +3\left({\dot a \over a}\right)\rho_{\rm h} +
            \frac{1}{a}\nabla \rho_{\rm h} \vec v_{\rm h} &=&
             \phantom{+}  \frac{\beta A \rho_{\rm c}}{t_*} 
                   -\frac{C\cdot\Lambda_r(\rho_{\rm h},T_{\rm h})
             }{
               \gamma\epsilon_{\rm h} -\epsilon_{\rm c}},\\  
\frac{\partial \vec v_{\rm h} }{ \partial t}
      +\frac{1}{a}(\vec v_{\rm h}\cdot\nabla)\vec v_{\rm h}
      +\left({\dot a \over a}\right)\vec v_{\rm h} &=&
         - \frac{1}{\rho_{\rm h}a}\nabla P_{\rm h} 
      - \frac{1}{a}\nabla\phi,\\
  {\partial E_{\rm h}\over \partial t}
    + 2\left({\dot a\over a}\right)E_{\rm h}
       + \frac{1}{a}\nabla\left((E_{\rm h}+P_{\rm h})\vec v_{\rm h}\right)
         &= &
 -  \frac{\gamma\epsilon_{\rm h} C\cdot\Lambda_r(\rho_{\rm h},T_{\rm h})+
       }{ \gamma\epsilon_{\rm h}
     -\epsilon_{\rm c}} -\Lambda_{\rm Comp}\nonumber  \\
          &\hfill{}& +  {\beta\rho_{\rm c} \over t_*}
      [\epsilon_{\rm SN} +A\epsilon_{\rm c}],
\end{eqnarray}
where $E_{\rm h}$ is the total (thermal plus kinetic) energy of the
hot gas per unit comoving volume ($E = \rho\epsilon+\rho v^2/2$), 
$kT_{\rm h}=(\gamma-1)m_H\mu_m\epsilon_{\rm h}$, $m_H$ is the
mass of the hydrogen 
atom, $\mu_m$ is the molecular weight per particle, $k$ is
Boltzmann's constant, and  
$P_{\rm h}=(\gamma-1)\rho_{\rm h}\epsilon_{\rm h}$

\item The Poisson equation:
\begin{equation}
  \Delta\phi = 4\pi G a^2 (\rho_{\rm dm}+\rho_*+\rho_{\rm gas} -
      \rho_{\rm b}), 
\end{equation}
\end{enumerate}
where $\rho_{\rm b}$ is the mean density of the Universe.

The collisionless dark matter (dm) and ``star'' ($*$) components
satisfy their continuity equations automatically by virtue of the
particle-mesh method.  Note that the pressure $P_{\rm disp}$
associated with the velocity dispersion of the cold clouds has been
implicitly included in Eq. (15), since $P_{\rm disp}\propto
\rho_cT_h$, $P_h\propto \rho_hT_h$, and $\nabla P_{\rm tot}/\rho_{\rm
tot} =\nabla P_h/\rho_h$.  As noted earlier, the contribution of cold
clouds to the density and the pressure turn out to be only a few per
cent.

The above system of equations is solved in two different regimes: If
the temperature of a volume element satisfies $T_{\rm h} > 2\times
10^5$K and if $\rho_{\rm c}$ is zero, then we suppose that the element
has not yet participated in star formation.  In this case the constant
$C$ is set to unity, and primordial chemical composition is
assumed. If either the temperature is lower or $\rho_c$ is not zero,
or there is already a significant star density  present, then solar
abundance is assumed with the cooling rate given by Raymond, Cox, and
Smith (1976).  Compton cooling is always included. We plan to make the
cooling rates more realistic by adding to the model the density of
heavy elements produced by supernovae and by modeling the effects of
photoionization in a self-consistent way as discussed in the
conclusions.

\section{Numerical techniques}
\subsection{Equations in expanding rescaled coordinates}

In order to apply the PPM method (Colella \& Woodward 1984) for gas
dynamics in a cosmological framework, we seek transformations from the
above physical quantities to rescaled quantities satisfying equations
that have the ``usual'' form of equations of gas dynamics (i.e., in
the absence of an expanding background). We adapted the transformation
used by Shandarin (1980).  First, we change to expanding coordinates.
Auxiliary dependent variables (denoted by a tilde) are related to the
physical variables as follows.  Density: $ \tilde \rho =a^3 \rho$;
peculiar velocity: $ \tilde {\vec v} = a \vec v$; pressure $\tilde P =
a^5 P$; internal energy per unit volume $ \tilde u = a^5 u$; internal
energy per unit mass $ \tilde \epsilon = a^2\epsilon$.  Second, we
introduce a new time variable
\begin{equation}
 b(t) = {2\over H_0}\left(1-a^{-1/2}(t)\right),
\end{equation}
where $H_0$ is the Hubble constant now.
For convenience, $b(t)$ was chosen to vanish at redshift zero ($a=1$).
Thus, $b$ is negative for $a<1$.
(In a nonflat background, the  formula for $b$ would be changed; the
definition of $b$ would be $da =\dot a a^2 db$.)

In terms of the new rescaled variables, the Poisson equation and the
Euler equation take the form:
\begin{equation}
\nabla^2 \tilde \phi =
        \frac{4 \pi G}{a}  (\tilde\rho_{\rm gas}+\tilde
   \rho_{\rm dm}+\tilde\rho_*-\tilde \rho_{\rm b}) ,
\end{equation}
\begin{equation}
{\partial \tilde{\vec v_{\rm h}} \over \partial b} +
 (\tilde{\vec v_{\rm h}}\cdot\nabla)\tilde{\vec v_{\rm h}} =
     - \frac{1}{\tilde \rho_{\rm h}} \vec \nabla \tilde P_{\rm h}
- a^2\vec \nabla \phi .
\end{equation}
The left-hand side of the continuity equation is transformed
back to the usual form
(i.e., no terms containing $a$),
but the terms on the right-hand side are multiplied by
a factor $a^5$. The same thing happens to
the energy equation: the left-hand side
takes the usual form, while the terms on the right-hand side 
are multiplied by a power
of the expansion parameter ($a^7$). In rescaled variables, the energy
equation and continuity equations for the hot gas, for the gas
component, and for ``stars'' take the form:
\begin{eqnarray}
{\partial \tilde E_{\rm h}\over \partial b}
 + \nabla\left((\tilde E_{\rm h}+\tilde P_{\rm h})
   \tilde{\vec v}_{\rm h}\right)  = &
    -  & {\gamma\tilde\epsilon_{\rm h}\over \gamma\tilde
   \epsilon_{\rm h}-\tilde\epsilon_{\rm c}} 
 a{C\tilde\rho_{\rm h}^2 L(T_{\rm h})\over (\mu_H m_H)^2}
 - D_{\rm Comp}\tilde\rho_{\rm h} T_{\rm h} + \nonumber \\
\hfill{} & +&  a^2{\beta\tilde\rho_{\rm c} \over t_*}(\tilde\epsilon_{\rm SN}
           + A\tilde\epsilon_{\rm c}) \\
 {\partial \tilde\rho_{\rm h} \over \partial b}  +
          \nabla \tilde\rho_{\rm h} \tilde{\vec v}_{\rm h}  =
     & \phantom{+}& a^2\frac{\beta A \rho_{\rm c} }{ t_*}
    - \frac{aC\tilde\rho_{\rm h}^2 L(T_{\rm h})/(\mu_H m_H)^2
   }{
      \gamma\tilde\epsilon_{\rm h}-\tilde\epsilon_{\rm c}} \\
{\partial \tilde\rho_{\rm gas} \over \partial b}  +
          \nabla \tilde\rho_{\rm gas} \tilde{\vec v}_{\rm h}  =
    & - & a^2\frac{(1-\beta) \tilde\rho_{\rm c} }{ t_*}\ \\
{\partial \tilde\rho_* \over \partial b}  +
          \nabla \tilde\rho_* \tilde{\vec v_*}  =
   &  \phantom{+} & a^2\frac{(1-\beta) \tilde\rho_{\rm c} }{ t_*}\\
\end{eqnarray}
Here we have defined $\Lambda_r(\rho_{\rm h},T_{\rm h}) =(\rho_{\rm
h}/\mu_H m_H)^2 L(T_{\rm h})$, where $\mu_H$ is the molecular weight
per hydrogen atom and $m_H$ is the mass of the hydrogen atom, and
$D_{\rm Comp}=7\times 10^{-36}/(\mu_H m_H)$.

\subsection{Solution of the hydrodynamical equations}
To calculate the flow of the ordinary (col\-lisional) matter,
described here as a multi-com\-ponent medium, we apply a
hydrodynamical code based on the PPM technique. One of the main
advantages of this Eulerian code is that it does not involve an
artificial viscosity, but still treats shocks very accurately (for
example, it does not generate post-shock fluctuations, and shocks are
only one zone thick). The method is very fast and highly
parallelizable.  This algorithm is applied to our gas-dynamic
equations in one dimension at a time.  Multiple dimensions are treated
by directional timestep splitting, whereas local processes (heating,
cooling, star formation, etc.), which involve various components of
collisional matter, as well as self-gravitation and gravitational
interaction with dark matter, are treated using process timestep
splitting (Oran \& Boris 1986).  Composition variables (such as the
concentration of heavy elements or the density of the gas $\rho_{\rm
gas}$) are advected through the mesh by the same PPM algorithm.

The time-scale of the cooling and star-formation processes usually
is shorter than the dynamical time-scale. In our code the time step
for advection of the gas and 
motion of the dark matter is equal for all zones, but cooling and
star-formation processes are integrated with a timestep adapted
for each zone.

\section{Tests and parameter studies}\label{tests}

\subsection{Hydrodynamic tests}

In order to check the accuracy and efficiency of the present
implementation, we first verified proper behavior in a number of
classical tests, two of which are presented here: Fig.~ 1 shows the
density and pressure for the Sod (1978) problem for the decay of a
discontinuity with the parameters $\rho_{\rm L} = 1$, $P_{\rm L} = 1$,
$u_{\rm L}=0$, $\rho_{\rm R} = 0.125$, $P_{\rm R} = 0.1$, $u_{\rm
R}=0$, $\gamma=1.4$.  All the important features of the flow (the
rarefaction wave on the left, the contact discontinuity in the middle,
the shock on the right) are well represented.  Fluctuations of
physical parameters in the vicinity of the shock and contact
discontinuities do not occur.  The shock and the contact discontinuity
are spread over only one cell.  These are important desirable
properties of the code.

In Fig.~2, we present the results of a strong ``2-D" cylindrical
explosion in an ideal gas with $\gamma=5/3$ as performed on 100x100
mesh with periodic boundary conditions.  The explosion energy is
initially deposited inside a small cylinder of radius 2.5 cell widths
located coaxially to the $z$-axis with $P=10^4$.  The pressure outside
the small cylinder is initially 0.1.  The initial  density is
$\rho=1$ everywhere.  The lower panels show the density at different
moments.  In particular, they demonstrate that oblique shocks are
stable.  In the left top panel, the density of every cell in the
simulation is plotted against its distance from the center of the
explosion.  The narrowness of the curves demonstrates that the shock
propagates in all directions at the same rate, and that the thickness
of the shock stays $\approx 1$ cell even after significant evolution.
The top right panel shows how the shock expands.  Note that in the
case of cylindrical symmetry the appropriate dimensionless variable is
$\xi=r \times (\sigma/(Et^2))^{-1/4}$, where $\sigma$ is the surface
density, rather than the usual "Sedov''-dimensionless variable $\xi=
r\times (\rho/Et^2)^{-1/5}$.  The resulting expansion rate is,
according to the Sedov solution, $r \propto t^{1/2}$.

\subsection{Supernova feedback and cooling enhancement parameters}
\label{s:test}

The evolution of the different baryonic components provides
information useful in understanding the role of the supernova feedback
parameter $A$ and the cooling enhancement factor $C$ as control
parameters in our nonlinear star-gas model.  Cooling and star
formation were studied in a single cell, i.e., with gravitation and
mass flow switched off.  Fig.  3 shows the time evolution of the
relative mass fraction of the different phases for models with varying
supernova feedback parameters $A=1$ (no evaporation), $A=20$ ($\approx
10$ per cent of supernovae energy goes to evaporation) and $A=100$
($\approx 50$ per cent goes to evaporation).  The gas was assumed to
have solar composition and initial temperature $T=10^6$K and density
$n_H=6\times 10^{-4}{\rm cm}^{-3}$, typical of a galaxy halo.  The
system was assumed to be in the regime of thermal instability with
cooling enhancement factor $C=10$, which as explained above accounts
for unresolved substructure in the interstellar medium.  Independent
of $A$, the hot gas initially cools with a characteristic cooling-time
of $(2-3) \times 10^7~\yr$.  At $t\approx 5 \times 10^6~\yr$, the
first stars form and begin to produce supernovae.  The supernovae
evaporate cold clouds, which is evident for example in the case $A=20$
as a rapid drop in the density of the cold gas and as a jump in the
density of the hot gas.  Although at early epochs ($t< 10^8$yr) larger
feedback parameter $A$ results in a smaller fraction of mass in cold
clouds and in ``stars'', the trend later reverses: high $A$ ultimately
causes more of the mass to end up in stars during the time available,
because the temperature of the hot gas stays lower and it cools
faster.  The star-formation e-folding time was $4\times 10^8~\yr$ for
$A=100$, $10^9~\yr$ for $A=20$ and $3\times 10^9~\yr$ for $A=1$.
Thus, due to the nonlinear feedback the
true star formation time scale is greater than $t_*=10^8~\yr$.

The results of this and other parameter studies illustrate that the
supernova feedback parameter $A$ is directly related to the efficiency
of star formation.  Indeed, in the context of a full simulation, the
effect of $A$ may be even stronger than implied by these studies:
Relatively larger values of $A$ imply that a larger fraction of the
energy supplied by supernovae will convert gas from the cold to the
hot phase.  As seen in Fig. 3, on the short term this conversion will
temporarily inhibit star formation, but by causing a smaller pressure
gradient it will allow at least the larger galaxies to keep most of
their gas.  On the other hand, relatively small values of $A$ imply
that a large fraction of energy supplied to the gas by supernovae will
heat the hot phase to even higher temperatures locally.  The ensuing
pressure gradient will enhance the expulsion of gas, even from large
galaxies, and retard star formation for a time comparable to a
dynamical time scale.  It is also clear that the dynamics of star
formation will be sensitive to the effects of environment such as
mergers, gas flow, tidal forces, etc., because these effects strongly
influence the amount and temperature of gas available for star
formation.

The value of $A$ should be chosen to allow for such phenomena as
``young, active galaxies'' and the transport of metal-rich, hot gas
from galaxies to the intergalactic medium.  
Because we are assuming $A$ constant in 
time, our choice will represent 
a compromise between the necessity of allowing for really active
regions of star formation (low values of $A$) and low efficiency of
conversion of gas to stars inside cold clouds (high $A$). 
These constraints suggest a value for
$A$ between 20 and 200.  
Another constraint comes from the condition that the total mass in
luminous matter (``stars'') be comparable with the observed value of
1\%-2\% of the critical density of the Universe. From our simulations we
infer that choosing the
feedback parameter in the range $A=100-200$ gives about the correct amount
of mass in ``stars''.

\section{Simulation of the effects of feedback mechanisms on galaxy formation}
   
\subsection{Cosmological model and initial conditions}

The initial fluctuations were computed as a realization of an
expanding flat CDM model (according to the approximation of Bardeen
\etal 1986) normalized to the rms mass fluctuation in a sphere of
radius $8h^{-1}$Mpc to be $\sigma_8 = 0.67$ as estimated by the linear
theory.  The ratio of baryonic to total density was taken to be
$\Omega_b=0.08$. 

We present results of ten numerical simulations whose parameters are
given in Table 1.  We used 128$^3$ particles and $128^3-256^3$ cells
in the simulations reported here.  The simulations were carried out on
SGI Power Challenge parallel supercomputers at NCSA and at the Centro
Europeo de Parallelismo de Barcelona (CEPBA).  About 700-1100
timesteps are needed to evolve a simulation to $z=0$ with timesteps
chosen to be 0.6 of the maximum according to the
Courant-Friedrich-Levy condition.  Due to the high degree of
parallelization of the code, one time step requires only about one
minute of real time on the SGI Power Challenge with 4 90MHz R8000
processors; it takes about 100 node-hours to run a simulation of 1000
steps.  For a 256$^3$ simulation, one time step takes
about 8 times as long, and 1 GByte of main memory is required.

The simulations were performed for a variety of 
different parameters and 
resolutions.  The three last lines of Table 1 give the fraction of mass
$\Omega_{\rm luminous}$ converted to stars and the redshift $Z_{1/2}$ at
which $\Omega_{\rm luminous}$ was half of its value at redshift zero.
 (We assume $H_0= 50 \kmsmpc$ throughout this paper unless otherwise stated.) 

The choice of box size clearly involves a compromise among different
considerations.  Sufficiently small cell size was the overriding
consideration for this paper.  The box size of 5 Mpc actually chosen
for Models 1-5 implies a cell size of 39~kpc, which is just large
enough to resolve galactic scales of interest.  At the same time, a
box of this size contains quite a few galaxies of different masses at
$z=0$.  It also typically contains a filament and thus incorporates at
least this aspect of the observed large-scale structure.  Models
1,3,4, and 5 were run with exactly the same initial conditions.  Model
2 had a different realization of initial fluctuations, but otherwise
the same set of parameters as Model 1. Model 10 was the same
realization as Model 2 (same box size and phases), but with 8 times as
many cells (size 19.5 kpc).

The smaller the box size, the greater the influence of the boundary
conditions:  The size of the largest structures in the box should be small
compared to the box, because mirror images due to the cyclically symmetric
Green's function would otherwise distort the results.  Also, a smaller
volume clearly reduces the number of galaxies to be expected at the end of
the simulation.  Nevertheless, in order to check the sensitivity of our
results to the effects of resolution, we ran another series of models
(Models 6-9) with comoving cell size 11.7~kpc (box 1.5 Mpc).  

Models 1--6 were evolved up to $z=0$, while  Models 7-10 were
evolved until $z=1$.  

\subsection{Simulation parameters and constants}
We now discuss and summarize the specific parameter values used here.  

The supernova mass fraction $\beta$ defined in \S 2.5 was computed
assuming the Salpeter 
initial mass function $dN_*/d(\log{M})\propto M^{-1.35}$ for masses
between $0.1\Msun$ and $100\Msun$. Assuming that stars with 
$M>10 \Msun$ explode as supernovae, we obtain $\beta=0.12$, the
value adopted here. 
One can consider the effect of steepening the IMF.  For example, 
an exponent of -1.6 would lead
to $\beta=0.05$.   Note that at the low-mass end, 
there is evidence for a decrease in the slope of the IMF 
(Miller\& Scalo 1979).  Thus, taking the lower mass limit
to $0.5\Msun$, one would obtain $\beta = 0.13$.  

In order to arrive at a value of the 
characteristic star formation time constant introduced above
$t_*=10^8$~yr, we note that i) even massive stars
live for $t_{\rm MS}=\allowbreak 3\times10^7[M_*/10\Msun]^{-1.6}\yr
\allowbreak \approx (1-3)\times10^7\yr$ before they explode as
supernovae, and ii) star-formation occurs in different clouds; it is
reasonable to suppose that the production of stars inside different
clouds in one simulation cell is not synchronous.  

In introducing a single constant $t_*$, it is important to recall that in
our picture, star formation is regulated by the availability of cold gas
clouds.  In codes that do not provide an independent estimate of the cold
gas component, a reasonable procedure is to estimate a local dynamical or
cooling time of the gas component in order to infer the amount of cold gas
available (e.g.  Katz, 1992; Navarro \& White 1993).  In our code, the
dependence of the star formation rate on the cooling time is already
implicit in the dynamics of the multiphase model.  

In \S \ref{tests}, we saw that the star formation rate depends strongly on
the supernova feedback parameter $A$.  In Models 1 -7 and 10, we take $A=200$,
corresponding to very efficient star formation and for comparison $A=100$
in Models 8-9.

``Stars'', represented by particles, were assigned a mass implied by
Eq.  (\ref{eq:srate}).  Once produced, a ``star'' particle moves like
a collisionless particle.  (In fact, these particles are best thought
of as representing starbursts.)  As explained in \S \ref{sect:22},
photoionization will tend to prevent cooling in low-density regions.
To incorporate this effect, a threshold of ${\cal D}=50 - 100 $  times the
background baryon density (see Table 1) was imposed for converting hot
gas to cold clouds in a cell (i.e., switching on star formation in a
cell).  This prescription also has the advantage of reducing the
computational cost of excessive low-mass starbursts.

As explained in \S \ref{s:modeleq}, solar abundances were assumed if either
star-formation conditions were met or stars were already present in a cell.
Now, the presence of formed stars in a volume element is an important
indicator that the gas was enriched by metals.  However, in simulations
reported earlier (Klypin, Kates, \& Khokhlov, 1992; Yepes et al., 1995 ,
1996a), the cooling regime was determined from star-formation criteria
alone.  Note that in these earlier simulations the hydrodynamic module was
based on the method of flux corrected transport (Boris \& Book, 1973,
1976).  Hence, as an additional
control, it was useful to run a few models (5,8, and 9, marked by asterisks)
without taking into account the presence of star particles when choosing the
cooling regime (primordial/solar composition). 
These runs also illustrate the sensitivity of results to the chemical
composition.  It turns out that the effect of taking primordial composition
instead of solar hardly affects the early evolution of Models 5, 8, and 9
compared to the others:  Since the temperature of the gas is still
comparatively low, the gas cools fast with either composition, producing
cold clouds. It also has little effect in small
galaxies at any epoch, because in small galaxies, winds (pressure gradients)
expel the heated gas in either case. But the evolution at later times ($z<1$) is different.  The
temperature of the gas in large galaxies grows with time because of the
energy released by supernovae and because more massive galaxies are capable
of holding on to the hot gas.  Thus, the cooling time
gradually increases in massive galaxies.
Now the system becomes sensitive to the assumed chemical composition. Eventually, the density of cold gas will drop below the threshold -- a few per
cent of the total gas density -- needed for a star formation to occur.  At
this moment, the models in which the local presence of stars is not taken
into account switch to primordial cooling, which effectively shuts off star
formation for the galaxy after $z=0.5$. If the local
presence of stars is taken into account, the code  is
more likely to switch to solar cooling and hence higher cooling rates.
Thus, the galaxy tends to produce cold clouds and new stars at a steady
rate.  In spite of this qualitative difference, which dramatically affects
star formation rates and thus the colors of large galaxies, the overall
effect on the total mass in stars and on the epoch of star formation
$Z_{1/2}$ was still not that significant (compare models 4 and 5 in Table
1).

\subsection{Galaxy identification algorithm} \label{sec:53}

Since the code produces separate estimates of all four density components
(dark matter, hot gas, cold gas, and stars), the information contained
in these four density fields could in principle be used in a variety
of ways to define ``galaxies''.  From the point of view of comparison
with observation, one is most interested in the distribution of 
visible matter, suggesting that the star density should be viewed
as the most important quantity to be analyzed in finding galaxies.  
Nevertheless, identification of dark matter halos proved
to be the most useful procedure for the purposes of this
study: In order to study the relationship between galactic 
 morphology (e.g., luminosity, spectral properties) and location
or environment, we obviously need to include low luminosity 
but relatively massive halos in the analysis.  

For this reason, we begin by constructing the total density $\rho$ on
the grid and searching for local maxima in $\rho$ such that $\delta
\rho \equiv \rho-\rho_b$ exceeds a fixed threshold density $\rho_{\rm
thresh}$.  (Here we took $\rho_{\rm thresh}=100 \rho_{\rm crit}$.)
Next, we place spheres of $r_{sp}=2$ cell radii around maxima and
determine the center of mass (CM) of the particle distribution (dark
matter and stars) within the spheres.  (In view of limited resolution
and the low mass fraction of gas, inclusion of the gas density would
have little effect.) We then translate the center of the sphere to the
CM and compute the mass $M_{sp}$ and the overdensity $(\delta \rho /
\rho)_{sp}$ within the sphere.  If $(\delta \rho / \rho)_{sp} > 200$,
we add the halo to our catalog.  If $(\delta \rho / \rho)_{sp} < 200$,
we reduce the radius to $r_{sp} =1$ cell radius.  If $(\delta \rho /
\rho)_{sp} > 200 $ in this sphere, then we add the halo to our
catalog; otherwise we do not include the halo.

For models with cell size 39~kpc, the lower galaxy mass
limit for $r_{sp}=2$ is roughly $2 \times 10^{10} \Msun$, while for
$r_{sp}=1$ the limit is roughly $2.5 \times 10^9 \Msun$.  We have
checked that no overlapping halos are identified with this algorithm.

\subsection{Stellar Population Synthesis}\label{ss:sps}

We wish to estimate the spectral energy distribution (SED) 
$S(\lambda,t)$ at an arbitrary 
epoch for each halo (referred to here as ``galaxy'') in the catalog.
The SED will then allow us to compute luminosities  and
 colors, to be analyzed below.
For each galaxy, we have 
a list of particles representing the stellar component.
Associated with each particle is the mass and time of formation 
of the ``starburst.''  Consistent with our assumptions
of constant IMF and star formation under
conditions of solar chemical abundances,
these two parameters allow us to compute 
the evolution of the SED from a stellar population synthesis model.

A number of stellar population synthesis models 
have been proposed (e.g. Tinsley 1972, Bruzual 1983, Guiderdoni \&
Rocca-Volmerange 1987, 1990, Buzzoni 1989, Charlot \&
Bruzual 1991, Bruzual \& Charlot 1993).  Among them, we have chosen the model
of Bruzual \& Charlot (1993) that describes the time evolution
 (between 0 and 20 Gyr)  of SED's for a burst of star formation
 under conditions of solar metallicity and 
a Salpeter IMF, in accordance with our model.
It covers the relevant
stages of stellar evolution, including the asymptotic giant branch (AGB)
and post-AGB stars, and it is based on
up-to-date stellar evolution calculations.

For each galaxy in the catalog, we may compute  
$S(\lambda,t)$ from
\begin{equation}
S(\lambda,t) = \sum_{\tau_i}  \Phi(\tau_i) {\cal F} (\lambda,t-\tau_i) , 
\end{equation}
where  $\Phi(\tau_i)$ is the mass of stars in the halo  produced at timestep 
 $\tau_i$  and 
${\cal F}(\lambda, t)$ is the SED of a starburst of 1 $\Msun$ 
after a time $t$.
Then we simply convolve $S(\lambda,t)$ with the filter response function
$R_f(\lambda)$ to obtain the
absolute luminosity $L_f(t)$ in  the given band.
Combining the $L_f(t)$, we then obtain the evolution of
the color index of the galaxy.  Of course, the color
of a galaxy that would actually be measured will be influenced
by other factors such as the interaction of starlight 
with the surrounding plasma.  

\section{Results}

We begin our discussion of the results by comparing some global
statistics concerning observational quantities with estimates obtained
from galaxy catalogs.  For example, in Table~2 we give the average
B-band luminosity density for Models $1 - 5$ (5 Mpc box). These values
were obtained by integrating all the light from the stars at $z=0$.
Depending on the different parameters, the luminosity density ranges
from $7-11\times 10^7 L_\odot/{\rm Mpc}^3$.  Observational estimates from
the APM redshift sample (Loveday et al., 1992), which covers the
largest volume surveyed to date, give a B-band luminosity density of
$6.2\times 10^7L_\odot/{\rm Mpc}^3$.  Taking into account that there could
be systematic photometric errors in their estimates (e.g. Ellis et
al. 1996; Bertin \& Dennefeld, 1996), a factor of 2 increase cannot be
ruled out.  This would imply that the amount of light produced in our
models is not far from what is estimated for the Universe.  Bolometric
mass-to-light ratios $M/L$ can also be computed from our simulations.
They range from 24 to 85 $M_\odot/L_\odot$ when averaged over a volume
of 5 Mpc on a side.  On the other hand, the star formation rate at low
redshifts in our simulations was a factor of $\approx 4.5$ higher than
the most recent estimate of $0.013 \Msun/{\rm yr}/{\rm Mpc}^3$ (Gallego et al.
1995) for our local Universe.

An important characteristic can be deduced from Table 2.  Although as
we shall see below the observational properties of individual halos
can be strongly affected by different values of the ``chemical"
parameters, the global statistics of the system are less sensitive to
these values.  Figs.  4 and 5 compare Models 1 and 2, respectively,
which have the same chemical parameters but different realizations of
the initial fluctuation spectrum.  In Figs.  4a and 5a, 3D views of
the dark matter distribution in the respective simulation boxes at
redshift $z=0$ are shown.  In Figs.  4b and 5b, the distribution of
``star'' particles is illustrated.  (Note however that the figures do
not take into account the different masses assigned to different
``star'' particles.)  Comparison of the figures reveals that there are
many dark halos, especially smaller halos, which have not succeeded in
producing stars.  ``Biasing'' is suggested by visual inspection of
these figures, in that the more massive objects contain a
disproportionate number of star particles.  The star concentrations
trace all of the large dark matter halos, but there is more variation
in the luminosity of smaller-mass halos.  There is a striking visual
impression that virtually all halos located in voids are completely
dark.

Figs.  6a (Model 1), 6b (Model 6), and 6c (Model 10) show contour plots of the
temperature distribution $T_{\rm gas}$ as well as the densities of luminous
matter (``stars''), dark matter $\rho_{\rm dm}$, and gas $\rho_{\rm gas}$ in a
thin slice of thickness 1 cell width passing through the centers of the most
massive galaxies in the simulations.  The filamentary structure is again
prominent in the gas distribution.  The most massive galaxy in Fig.  6a (just
above the center) is rather large -- it has mass $7\times 10^{11}\Msun$ and 3D
rms velocity 250~km/s.  The internal gas temperature is $2\times 10^6$K.  The
galaxy is surrounded by a large (1/2~Mpc radius), low-density ``cavern" of hot
$\sim 10^6$K gas.  The cavern is produced by stellar winds blowing from the
galaxy and by a converging flow of colder gas falling onto that galaxy,
primarily from the filaments. The hot gas tends to expand into low-density
regions.  

There are two small satellites of the galaxy inside the cavern.  In spite of
their small mass, these satellites have managed to produce a significant amount
of luminous matter as clearly seen in the upper right panel of Fig 6a.  Compare
those small galaxies with their counterparts further from the large galaxy
(e.g.  at $x=3.5, y=2$, or at $x=2, y=4.7$), which do not produce ``stars''.
This is an example of biasing depending on the environment.

Galaxies in Fig. 6b (Model 6 at $z=1$) are significantly smaller than
the large galaxy in Fig. 6a.  The largest galaxy in Fig. 6b (at the
center) has 3D rms velocity 80~km/s. As the result, the size of the
hot gas cavern and the temperature of its gas are significantly
lower. Nevertheless, the basic pattern is similar: star formation
predominates in large galaxies and in smaller galaxies which are inside 
relatively dense filaments. Small dark halos outside the filaments
(and far from a large galaxy) do not generate ``stars'' and
luminosity.

In Fig. 6c, which represents the high resolution Model 10 at $z \approx 1$,
the pattern of preferential formation of stars along the filament
continues, even for dark matter halos with a significant amount
of gas.  The change in resolution does not lead
to a qualitative change in the picture.
 
There is a strong statistical correlation between the density of the
luminous matter and the density of the dark matter. In Fig. 7a the
density of the luminous matter and the density of the dark matter are
plotted for every cell in the simulations. Points with density
contrast less than 200 represent either the periphery of galaxies or
lie entirely outside galaxies. They show very little correlation of
the luminous matter with the dark matter. But high density and high
luminosity density regions demonstrate quite a remarkable correlation,
which we estimate as
\begin{equation} \label{eq:dmcor}
   \rho_{\rm lum}=(\rho_{\rm dm}/133)^{1.7},
\end{equation}
for $z=0$ and approximately 3 times smaller amplitude for $z=1$. The
only models which had significant deviations from this correlation are
Models 8 and 9 (the latter is not shown in Fig. 7a). Those models were
run for $A=100$ -- one-half of the value in the other models. 
A smaller value
of the feedback parameter resulted in significant suppression of 
star formation in high-density regions. Those models had $\sim 3$
times smaller total amount of luminous mass (Table 1). It is
interesting to note that changing the cooling rates by ten times
(Model 1 vs. Model 3; Model 8 vs. Model 9) does not have much of an
effect. The scatter of this relation is still
significant -- a factor of 2-3. It leaves room for the non-local biases
observed in Figs. 5 and 6.

In Fig. 7b (Model 10), we observe (up to a constant of proportionality)
the same trend at redshifts
4, 2.3, and 0.9 as in Eq. (\ref{eq:dmcor}).  There is no effect
of resolution.  This provides evidence that the relation
could retain its validity at higher densities.

In Figs. 8,9, and 10 we plot different characteristics of ``galaxies''
identified with our algorithm (see \S \ref{sec:53}): the ratio of
luminous to total mass $M_{\rm stars}/M_{\rm tot}$ as a function of
the total mass of the galaxies (Fig. 8), the absolute visual magnitude
of the halos versus total mass (Fig. 9), and the UBV color-color
diagram (Fig. 10).  The different plotting symbols in the figures
indicate the distance to the most massive galaxy.  The dashed line in
Fig. 8 shows the average baryon fraction $\Omega_B=0.08$.  For
high-mass ``galaxies'' in Fig.8 ($M_{\rm tot}> 3\times 10^{10}\Msun$),
there is a trend in the mass fraction in luminous matter as a function
of the total mass, which we approximate as $M_{\rm stars}/M_{\rm
tot}\approx 0.035 \sqrt{M_{\rm tot}/10^{11}\Msun}$. The trend
indicates that larger galaxies are more efficient in converting gas
into stars.  It may be understood by imagining that the efficiency of
gas ejection due to supernovae is a strong function of the mass of the
galaxy, at least in some range of galaxy masses, because positive
pressure gradients, which are not primarily dependent on the mass of
the object, will be balanced by relatively large inward gravitational
forces, which {\it are} related to the mass of the object.  At the
very high-mass end, for $A=200$, gravitational forces already suffice
to avoid gas loss, and therefore further increases in mass would have
a less dramatic effect on star formation, leading to a flattening of
the curve.  For lower values of the feedback parameter $A$ (Models 8
and 9), the pressure gradients are higher because more energy from
supernovae goes into reheating the hot gas. The net result is a global
decrease in the amount of stars formed for some given total mass. This
effect is clearly seen in Fig. 7a for Model 8 (lower right
panel). (See also Yepes et al. 1995, 1996)

As the mass decreases, the $M_{\rm stars}/M_{\rm tot}$ curve first flattens off
and then develops a large scatter at low masses.  The dramatic variation in the
fraction of luminous matter down to zero is clearly an indication that the
gravitational potential of these galaxies was not deep enough to bind the gas
heated by the first supernovae which occurred at early times when the galaxies
first collapsed.  These are the galaxies that did not sustain even the first
burst of star formation.  Those low-mass galaxies with relatively high star
mass fractions and/or higher luminosities had their gas replenished by the
environment.

Slightly larger galaxies (masses around $10^{10}\Msun$) had another history.
We have traced the star formation rates of few of these galaxies (Figs.  11 and
12).  These galaxies were able to keep their hot gas for some time:  the
Universe was denser at high redshifts, and the gas was still cooling rather
fast and falling back onto the galaxies.  But as time went on, the gas was
getting farther away and its cooling time was getting longer.  This resulted in
a significant decline in star formation rates at small redshifts.  Because the
star formation rate peaked at high redshifts ($z=3-5$) for small galaxies,
those galaxies are relatively less luminous and redder at present.  Because of
that the flattening in the ratio $M_{\rm stars}/M_{\rm tot}$ at $M\sim
10^{10}\Msun$ (Fig.  8) is barely seen in the luminosity $M_V$ (Fig.  9).  It
is possible to check the effect of spatial resolution on the peak of star
formation by considering the higher resolution Models 6--10.  We find the same
trend:  the peak of the star formation rate tends to higher redshifts for
progressively less massive galaxies.

The dependence of the
absolute visual magnitude $M_V$ on the total mass can be approximated as 
\begin {equation} \label{eq:MV}
 M_V=-18.5 -4\log \left( \frac{M_{\rm tot}}{10^{11}\Msun} \right).
\end {equation}  
This relation fits all our models for $M_{\rm tot} > 2\times
10^{10}\Msun$ and $M_V < -15$ and has very small deviations of less than
1/2 magnitude. 

A large spread in the fraction of luminous matter is apparent for the
low-mass halos.   Figs. 6 give evidence that low-mass halos located
along filaments are more likely to have a higher mass fraction of
stars than those not located on or near the filament.
Such a correlation would be expected in a scenario in which gas traces the
local filamentary structure and is thus preferentially available for star
formation in those galaxies located along the filament.  As explained earlier,
for the value of the supernova feedback parameter $A=200$ used in most of the
models, the expulsion of gas from dense regions due to supernovae is relatively
inefficient compared to evaporation of cold clouds, increasing the amount of
hot gas at the expense of its temperature.  In massive galaxies, this gas
remains in the potential well, cools, and falls back into the central regions
to allow later epochs of star formation.  Now, it is true that even for $A=200$
supernovae would eject most of the gas out of the potential well of a small
galaxy, but there is a relatively large reservoir along the filament including
cold gas that can eventually lead to some future star formation.

Fig.  10 is a $UBV$ color diagram for galaxies brighter than $M_V<-15$
in our catalog (in the rest frame of the galaxies).  In each panel,
the horizontal and vertical axes represent the Johnson $B-V$ and $U-B$
color indices, respectively, computed as explained in \S \ref{ss:sps}.
The long dashed curve represents the $UBV$ color diagram for main
sequence stars of luminosity class V.  For reference, the short dashed
line represents fully corrected colors of observed galaxies in the
range from Hubble types +10 (irregular) at the upper left end to $-6$
(E0) at the bottom right end (de Vaucouleurs, 1977).  The typical
deviation is about 0.1 magnitude.

The simulated galaxies are roughly consistent with the observed
relation, although there is a slight blue trend.  The blue trend is
reduced in Model 4 (lower right) compared to Model 3 (lower left).  In
particular, note that in the lower left panel $B - V$ for the most
massive galaxy in the simulation is shifted by about 0.2 toward bluer
colors compared to the lower right panel.  The bluer color is
associated with a higher rate of star formation at recent times as
discussed for Fig.  12 below.  The color differences may be
attributable to the lower density threshold ${\cal D}$ of Model 4 for
allowing star formation.  We recall that the density threshold is
intended to incorporate in a simple way the effects of heating in
lower-density regions due to photoionization.

Note that, according to the Bruzual \& Charlot (1993) model used here, a $B-V$
color index greater than 1 as in elliptical galaxies would require more than 13
Gyr, and in view of $H_0=50\;\; \kmsmpc$ this possibility is therefore excluded.

In Fig.  11, the star formation rate (SFR) is plotted as a function of
redshift for three galaxies in Models 1 and 6: Star formation begins
very early ($z>9$) in the largest galaxy (solid line), and the SFR
rises up to a maximum of $\sim 15 \Msun/{\rm yr}$ at about 12 Gyrs ago
($z=2$), after which it falls steadily towards a present SFR of $(2-3)
\Msun/{\rm yr}$.  In the third-largest galaxy (dashed line), star
formation begins later, rising to a maximum at about $z=2.5$ and then
falling off more sharply.  Note that Model 6 has more than 3 times
better resolution than model 1 (5.8~kpc at $z=1$) and yet it basically
shows the same results: the peak in star formation is higher and it
occurs later for more massive galaxies.  The decline is stronger for
smaller galaxies, which means that dwarf galaxies (like those least
massive shown in this plot) are older and redder than more massive
galaxies.  The periodic bursts of star formation exhibited by the
small galaxy (dotted line) are rather typical for the dwarf galaxies
in our simulations.  This behavior is due to the shallow potential
wells of these halos, which are unable to retain the gas heated by
supernova explosions (Yepes et al 1995).  The positive pressure
gradient drives the hot gas out of the central region, but some of it
will still manage to remain in the outer parts of the potential well.
The gas may then cool, most efficiently in the central regions, and
subsequently recollapse, giving rise to a new burst of star formation.
In this sense, our simulations seem to lend support to some previous
analytical studies of the nature of dwarf galaxies in the Universe, in
which the periodic bursts of star formation for these objects and
their dependence on environment were predicted (Dekel \& Silk 1986;
Babul \& Rees 1992; Campos-Aguilar, Moles \& Masegosa 1993)

Fig. 12 illustrates the dependence of the SFR on model parameters
while holding the phase relationships in the initial realization of
the perturbation spectrum constant.  In the upper panel, the upper
curves are for the most massive galaxy, while the lower curves are for
the least massive luminous galaxy.  The solid lines correspond to
Model 1 (also seen in Fig.  11); the dotted lines are for Model 3 (no
enhanced cooling); while the dashed line is for Model 4 (lower
star-formation threshold).  The lower panel is for the high-resolution
simulations: the solid lines are for Model 6, the dotted lines for
Model 7 (no enhanced cooling), and the dashed lines are for Model 8
(lower supernova feedback parameter).  At high redshifts ($z>4$)
galaxies of a given mass category behave almost in the same way
independent of our simulation parameters.  At redshift one, Model 8
(smaller feedback parameter $A=100$) gives a factor of at least five
lower star formation rates than the models with $A=200$.  As pointed
out earlier, for $A=100$ a smaller fraction of the supernova energy is
used to transfer gas from the cold to the hot component, and the net
effect is to raise the temperature of the hot gas more, thus
increasing outward pressure gradients.  Strangely enough, Models 3 and
7 (no enhanced cooling) actually have a somewhat higher rate of star
formation at recent times as discussed in Fig. 10.  All in all, the
effect of changing $C$ by a factor of 10 is not as dramatic as one
might have suspected.

Above, we noted that the global star formation rate at recent times
seems to be higher than the observed rate even though the luminosity
comes closer to agreeing.  A separate but related issue is the even
higher star formation rate early in the simulations, which reflects
the prevalence of early objects due to high small-scale power in the
standard cold dark matter spectrum.  This is also reflected in the
very red colors of most of our old dwarf galaxies.

\section{Conclusions}

Confrontation of scenarios for large-scale structure formation with
observations require an assignment of observational properties such as
luminosity and colors to mass concentrations.  The procedure utilized
here provides such an assignment using a model based on astrophysical
processes in the interstellar plasma which (with reasonable
confidence) we believe must occur, even if the details are not
completely understood.  While improvements in numerical resolution are
always welcomed, at some point in any approach it will be necessary to
estimate the effects of processes occurring below the resolution on the
basis of what is known about quantities averaged over the resolution
scale.  These estimates demand careful consideration of the
small-scale nongravitational physics, which plays a major role in the
most important observable consequences.

All of these effects contribute to and complicate the relation between
the luminosity and colors of galaxies on total mass.  At sufficiently
small scales, the aforementioned strong coupling of morphology to
dynamics and thus to physical location poses severe problems for any
simple-minded biasing prescription such as those frequently used in
large-scale structure simulations.  This conclusion generally supports
our working hypothesis that { \em improvements in resolution resulting
from improved computing resources will only lead to better predictions
if the most important processes occurring at small scales are properly
modelled.  } As we saw above, the colors of the brightest galaxies
found here are dominated by young stars, tending to younger stellar
types with increasing luminosity.  However, it is too early for any
sweeping statement.

The results of this paper underline the rather complicated interaction
of gas dynamics, supernova feedback and gravitation in the evolution
of observable properties of galaxies.  Gravitational condensations are
of course a prerequisite in order that an adequate supply of cold gas
be available for star formation.  Subsequently, gravitation itself
plays more of a passive role: the most critical processes regulating
the dynamics of star formation (and ultimately the observable
properties of galaxies) are the various star-gas interactions and
feedback loops.  The most important of these is the supernova feedback
loop, which regulates star formation in three different ways, two
inhibitory and one stimulatory: First the evaporation of cold clouds
reduces the efficiency of star formation.  Second, the enormous
transfer of heat to the hot gas component results in pressure
gradients, which work against gravitational forces and expel gas,
especially from shallow potential wells. This effect strongly couples
subsequent star formation to hydrodynamics, and in particular provides
a possible mechanism for a strong dependence of morphology on
position.  Third, metal production due to supernovae further enhances
cooling.  This third effect is only seen locally in our code, because
composition gradients are not advected, but it should in principle
cause some additional further coupling of the hydrodynamic and
supernova feedback loops.

A relation between the absolute visual magnitude $M_V$ and the total
mass was given in (\ref{eq:MV}) for $M_{\rm tot} > 2\times
10^{10}\Msun$.  Some of the ``scatter" in this relation can presumably
be attributed to the environment, but much more work will be needed to
establish a pattern.

In this paper, effects of photoionization were included only in a
phenomenological way.  We would expect photoionization to inhibit star
formation most effectively at the low-mass end of the mass luminosity
relation.  There would be several important effects.  First, one might
expect relatively more young stars and therefore increased visual
magnitude now.  Second, the colors would be shifted toward the blue.
For larger galaxies, we would expect fewer modifications of the
present results.  However, it seems unlikely that proper inclusion of
these effects would cancel out the most important qualitative
conclusions of this paper, namely that there is a strong coupling
between gas dynamics and the supernova feedback loop, which in turn
act as a kind of ``hidden variables'' and influence the observable
properties of galaxies.  In particular, the dependence on ``hidden
variables'' would be expected to increase the scatter in relations
such as the mass-luminosity relation.  The scatter seen here decreases
strongly with mass.  However, one might suspect that if it were
possible to keep the same spatial resolution while working in a much
larger simulation box, the effects of large-scale structure would lead
to additional scatter at the high-mass end as well.  Definite
conclusions in this direction will unfortunately have to await further
improvements in computational resources.

Because the problem contains ``hidden variables," there is no way to
avoid uncertainties and scatter in the predictions based on averaged
quantities.  However, at least with regard to star formation, it is
possible to avoid or at least reduce {\it systematic} errors resulting
from not resolving small-scale densities by incorporating the
available information into a characteristic star formation timescale
$t_*$.  It can justifiably be argued that (for a given level of
computer resources) some other hydrodynamical method such as SPH could
achieve higher resolution in high-density regions than our methods.
(I.e., some of the small-scale information which is ``hidden" to us
would thus no longer be hidden.)  This better resolution has the
potential to produce a more accurate estimate of cooling.  Of course,
at still smaller scales cooling is complicated by processes such as
thermal instabilities, clouds, and the effects of a multicomponent
medium.  Nonetheless, it is still a valid question to ask how much
influence fluctuations in density on scales below the resolution of
our Models 1 - 10 could have.  Although it is too early for a
comprehensive answer, we can get a partial answer by comparing the
results of the simulations with cooling enhancement parameter $C=1$
(Models 3,7,9) with the $C=10$ cases.  The differences are not nearly
as striking as the dependence on the supernova feedback parameter $A$.
The reason may be that (enhanced) cooling is embedded in a feedback
loop which couples all the processes and their timescales.  If the
timescale for an effect is already the shortest in a loop, then it
makes little difference how short the timescale is.

Moreover, a good characterization of the diffuse, lower-density, but
hot gas is absolutely essential for a proper understanding of the
supernova feedback loop: First of all, some supernovae (about 3 \%)
are known to explode in the halos of spiral galaxies, so that the hot
gas density is estimated correctly by our code.  Second, even if a
supernova explodes within the disk of a galaxy, the resulting
supernova remnant will certainly deposit a significant proportion of
its energy in the hot interstellar gas.  Within the disk, which we do
not resolve, the typical density and temperature are roughly
$10^{-2.5}$~cm$^{-3}$ and $10^{5.7}$, respectively, and the cooling
time would be on the order of $10^6$~yr (McKee \& Ostriker, 1977),
which we underestimate somewhat even with enhanced cooling.  However,
a significant fraction of the supernova energy would escape to the
halo, where the cooling rate is either correctly estimated or even
somewhat overestimated by our prescription.  Hence, it is impossible
to argue that energy input due to supernovae is entirely locally
dissipated and thus irrelevant for the gas dynamics on larger scales
(Baron \& White 1987; Navarro \& White 1993; Kauffmann, White \&
Guiderdoni, 1993; Metzler \& Evrard, 1994)

In addition to the simulations reported in detail here, our experience
with a large number of test simulations shows that the star formation
history of galaxies and thus their observable properties such as
colors and luminosities may vary quite strongly from one simulation to
another depending crucially on the dynamics, especially interactions
producing shocks.  For example, if a galaxy merger just manages to
occur at low redshift (shortly before the end of the simulation), then
the ensuing burst of star formation dramatically increases the
luminosities of the affected galaxies and shifts them toward ``bluer"
color indices.  Hence, as a general statement concerning the
interpretation of simulations we conclude that there is a very rich
variety of interactions between large-scale and/or medium-scale
structure and galaxy evolution that will require many numerical
experiments and/or larger box sizes for an understanding.  The trend
toward increasing computer resources will allow larger boxes to be
achieved and will thus hopefully lead to a more general understanding
of statistical relationships between the environments of galaxies and
their observable properties.

{\em Acknowledgements:}
 
We are very grateful to G.  Bruzal for kindly making his models
available to us.  GY would like to thank C. S.Frenk, J.F.Navarro and
A.  Campos for very stimulating discussions.  REK is grateful to V.
M\"uller, J.  M\"ucket, and S.  Gottl\"ober for numerous productive
discussions and moral support.  GY also wishes to thank C.  S. Frenk
for inviting him to Durham.  The visit was financed by the European
Union Capital and Mobility Network {\em Ulysses}, whose support is
gratefully acknowledged.  This research was supported in part by NSF
grant AST-9319970 (AK) and DGICyT of Spain (GY, project PB90-0182).
REK was supported in part by a Fellowship (Ka1181/1-1) from the DFG
(Germany).

\def\reference{\bibitem[\protect \citename{}]{ciiaa}~}



\clearpage

\begin{table}
\caption{ Parameters of the simulations} 
\small
\begin{tabular}{ l c c c c c c c c c c} 
\multicolumn{1}{c}{Parameters} &\multicolumn{9}{c}{MODEL \# }\\
%
& 1   & 2   & 3  & 4  &5 * & 6  & 7  & 8*  & 9* & 10\\
Box Size (Mpc)                                 & 5   &5    &5   &5   &5   & 1.5& 1.5& 1.5 & 1.5 & 5\\
Formal resolution at $z=1$ (kpc)               & 19  &19   &19  &19  &19  & 5.8& 5.8& 5.8 & 5.8 & 9.5\\
Mass resolution for dark matter ($10^6M_{\odot}$)& 4 & 4   &4   &4   &4   & 0.1& 0.1& 0.1 & 0.1 & 4\\
Feedback parameter $A$ (eq.8)                  & 200 & 200 &200 &200 &200 & 200& 200& 100 & 100 & 200\\
Cooling enhancement factor $C$ (eq.3)          & 10  & 10  &1   &10  &10  & 10 & 1  & 10  & 1   & 10\\
Overdensity limit for star formation           & 100 & 100 &100 &50  &50  & 100& 100& 100 & 100 & 100\\
$\Omega_{\rm luminous}$ at $z=0$ (\%)          & 2.3 & 2.4 &2.1 &2.7 &2.4 & 2.1& --  & --   & --   & --\\
$\Omega_{\rm luminous}$ at $z=1$ (\%)          & 0.84& 0.77&0.77&1.4 &1.4 &0.68&0.68& 0.27&0.26 & 0.78\\
$Z_{1/2}$                                      & 0.65& 0.75&0.63&1.0 &1.2 &0.50& --  & --   & --   & --\\
 \end{tabular}
 \end{table}

\begin{table}
\caption{ Volume average observational quantities at  $Z=0$}
\small
\begin{tabular}{ l  c c c c c } 
\multicolumn{1}{c}{Quantity} & \multicolumn{5}{c}{ MODEL \#} \\
									                    & 1    &  2  & 3    & 4   & 5 \\
Luminosity density in B ($10^7 L_\odot/\Mpc^3$) & 9.4  & 8.0 & 11.0 & 8.5 & 5.8 \\
Bolometric $M/L_{\rm bol}\;\;(\Msun/L_\odot)$  & 31   & 31  & 24   & 28  & 85 \\
SFR density  ($\Msun/{\rm yr}/\Mpc^3)$            & 0.12 & 0.13& 0.11 &0.15 &0.13\\       
SFR density since $Z=0.05$($\Msun/{\rm yr}/\Mpc^3)$       &0.08 &0.07 &0.06 &0.09 & 0.08\\
Star density ($10^9\; \Msun/\Mpc^3)$            & 1.6  & 1.6 & 1.5  &1.9  &1.7\\
 \end{tabular}
 \end{table}

\clearpage

{\bf FIGURE CAPTIONS}

\begin{itemize}
\item [\bf Fig 1] The density (open circles) and the pressure (filled
   circles) for a one-dimensional shock tube.  The solid line represents
   the analytic Sod solution.

\item [\bf Fig 2] The evolution of the density for a strong ``2-D" cylindrical
explosion iexplosion in an ideal gas with $\gamma=5/3$.  A 3-D version of the
code on a $100\times 100\times 1$ grid was used with periodic boundary
conditions along each spatial direction.  The explosion energy is initially
deposited inside a small cylinder of radius 2.5 cell widths located coaxially to
the $z$-axis with $P=10^4$.  In the left top panel, the density of every cell in
the simulation is plotted against its distance from the center of the explosion.
The top right panel compares the shock expansion with the analytical Sedov
solution.  The lower panels show the density at different moments.

\item [\bf Fig 3] Effects of cooling and supernova feedback simulated in a
single cell.  No gravity and mass flow are taken into account (constant
density).  The time evolution of the mass fractions of cold gas, hot gas and
stars are shown for different values of the supernova feedback parameter $A$:
$A=1$ (dash - dot), $A=20$ (solid) and $A=100$ (dashed).  The gas, which has
solar composition and initial temperature $T=10^6$K and density $n_H=6\times
10^{-4}{\rm cm}^{-3}$ cools with enhancement factor $C=10$ (see text).

\item[{\bf Fig. 4}]

(a) A 3D view of the dark matter distribution in Model 1 at $z=0$. 
A random sample of 10\% of the dark matter particles is 
shown.  Dimensions are in Mpc ($H_0=50$ Km s$^{-1}$ Mpc$^{-1}$)

(b) Same as (a) but for the stars created in the simulation.
The points in this figure do not directly
represent the mass distribution of stars
because star particles have variable masses.

\item[{\bf Fig. 5}]

The same as Fig. 4, but for Model 2.

\item[{\bf Fig. 6}] 

(a) The distribution of temperature and densities in a slice of 1 cell
size (39~kpc) passing through the center of the brightest galaxy
of Model 1 at $z=0$.  Contours for quantities 
in parentheses are represented by thick solid curves. Lower contour
values are represented by
dashed curves. The density of
the dark matter is expressed in units of the critical density $\rho_{\rm
cr}$. The densities of the luminous matter (``stars'') and the hot gas are
expressed in units of the mean baryon density  $\Omega_b\rho_{\rm cr}$.

(b) Same as (a), but in this case for Model 6 at $z=1$.  A cross section of the
main filament running across the simulation box is clearly seen in the density
contours of dark matter, gas and stars.  The coordinate axes are scaled in
proper (physical) units.

(c) Same as (b) but for the high-resolution ($256^3$ grid cells) Model 10.   

\item[{\bf Fig.  7} ] Relation between the luminous and dark-matter
densities (both in units of the critical density).  The density of the
luminous matter and the density of the dark matter are plotted for
every cell in the simulations.  (a) Left panels are for Models 1,3, and 4
(bottom to top); Right panels are for Models 6,7, and 8 (top to
bottom).
(b) Same as (a) but for the high-resolution Model 10 at 
selected redshifts.

\item[{\bf Fig. 8} ] Fraction of luminous (star) mass to total mass as
 a function of the total mass for the galaxies in Models 1--4.
 Different symbols are used to plot galaxies at different relative
 distances to the most massive galaxy in the simulations. The dotted line
 represents the initial average baryon fraction (0.08).

\item[{\bf Fig. 9}  ]

Absolute visual magnitude of the galaxies in Models 1--4
 versus total mass. Symbols as in Fig. 8.

\item[ {\bf Fig.  10}] UBV color-color diagram for the bright galaxies $M_V<-15$
in Models 1 -- 4 (see figures).  The long dashed curve represents the UBV
position for main sequence stars of luminosity class V.  The short dashed line
is for galaxies (see text for details).

\item[{\bf Fig. 11}]

Star formation rates as a function of redshift for 3 different
galaxies in Model 1 (top panel) and Model 6 (bottom panel). The solid
line corresponds to the brightest galaxy in the respective simulation. 

\item[{\bf Fig. 12}]

Dependence of star formation rates on physical parameters.  In the
upper panel, the upper curves are for the most massive galaxy, while
the lower curves are for the least massive luminous galaxy.  The solid
lines correspond to Model 1 (also seen in Fig.  11), the dotted lines
are for Model 3 (no enhanced cooling), while the dashed line is for
Model 4 (lower star-formation threshold).  The lower panel is for
high-resolution simulations: the solid lines are for Model 6, the
dotted lines for Model 7 (no enhanced cooling), and the dashed lines
for Model 8 (lower supernova feedback parameter).

\end{itemize}

\end{document}